\documentclass[epjCONF,columns]{svjour}

\usepackage{graphics,epsfig,amsmath, amssymb}
\usepackage[varg]{txfonts} 
\usepackage[latin1]{inputenc}
\pdfoutput=1
\session-title {Hadron Collider Physics Symposium, 2011, Poster Session}

\begin{document}

\title{Track-Based Alignment of the Inner Detector of ATLAS}
\author{Ana Ovcharova\thanks{\email{Ana.Ovcharova@cern.ch}} on behalf of the ATLAS Collaboration}

\institute{Lawrence Berkeley National Laboratory (LBNL), USA}

\abstract{
ATLAS is a multipurpose experiment at the LHC. The tracking system of ATLAS, embedded in a $2~T$ solenoidal field, is composed of different technologies: silicon planar sensors (pixel and microstrips) and drift-tubes. The procedure used to align the ATLAS tracker and the results of the alignment using data recorded during 2010 and 2011 using LHC proton-proton collision runs at 7 TeV are presented. Validation of the alignment is performed by measuring the alignment observables as well as many other physics observables, notably resonance invariant masses in a wide mass range ($K_{S}$, $J/\Psi$ and $Z$). The $E/p$ distributions for electrons from $Z\rightarrow ee$ and $W\rightarrow e\nu$ are also extensively used. The results indicate that, after the alignment with real data, the attained precision of the alignment constants is approximately 5 $\mu m$. The systematic errors due to the alignment that may affect physics results are under study.
} 

\maketitle

\section{Introduction}
\label{intro}
\begin{figure}
\resizebox{0.9\columnwidth}{!}{\includegraphics{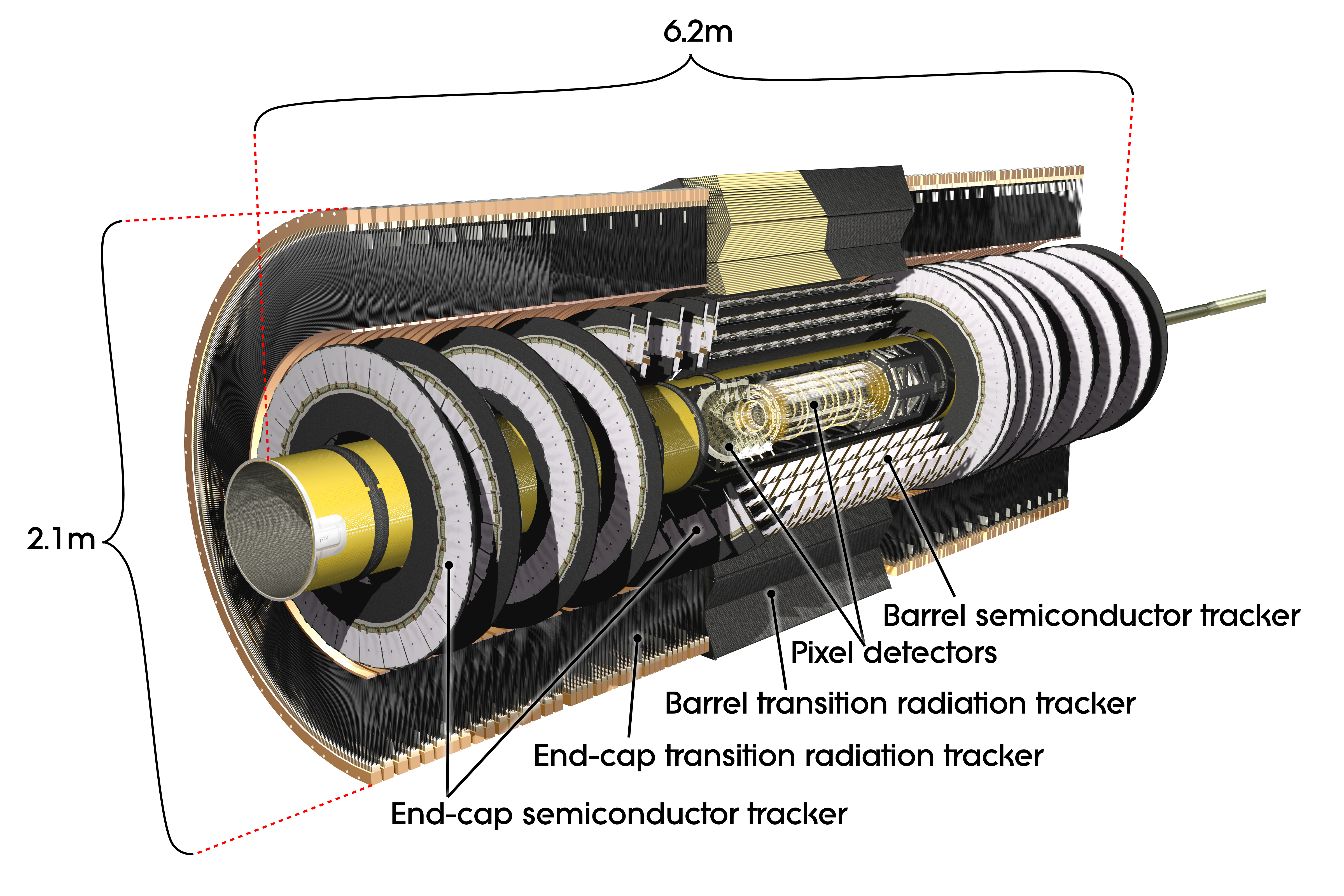}}
\caption{The ATLAS Inner Detector.}
\label{fig:1}       
\end{figure}

The ATLAS Inner Detector (ID), shown in Fig. \ref{fig:1}, is composed of the Pixel, the Semiconductor Tracker (SCT) and the Transition Radiation Tracker (TRT). The ID is designed to achieve the momentum and vertex resolutions required for high-precision measurements \cite{ATLASPaper}. To ensure that the resulting requirements on track reconstruction are met, the position and orientation of each active detector element must be known with sufficient accuracy such that track parameter resolution is degraded by less than 20\% of the design values. The following is an outline of the procedure, results and some of the challenges of the alignment of the ID.

\section{Alignment strategy}
\label{sec:1}
The alignment is derived by minimizing track residuals which are defined as the difference between the expected and the measured hit positions. The $\chi^{2}$ function to be minimized is given by:
\begin{equation*}
\chi^{2}=\sum_{Trks} \vec r(\boldsymbol{\vec\tau}, \vec a)^{T} V^{-1} \vec r(\boldsymbol{\vec\tau}, \vec a), 
\end{equation*}
where $V$ is the hit covariance matrix and $\vec r(\boldsymbol{\vec\tau}, \vec a)$ is the vector of track residuals, which are a function of both the track parameters, $\boldsymbol{\vec\tau}$, and the alignment constants, $\vec a$. ID alignment implements two flavors of $\chi^{2}$-based algorithms: the Global $\chi^{2}$ and the Local $\chi^{2}$. In the Global $\chi^{2}$ approach, a simultaneous minimization with respect to all track parameters and alignment constants is done. This approach ensures that full correlation between alignable objects intersected by a common track is retained. In the Local $\chi^{2}$ minimization, module correlations are discarded, rendering alignment less computationally intensive. However, it is necessary to perform multiple iterations to reach convergence. The alignment uses isolated high-$p_{T}$ tracks to reduce the impact of pattern recognition ambiguities and of multiple scattering. Both collision and cosmic ray tracks are used to maximize long-distance correlations between detector elements. ID alignment is staged at several levels of granularity, corresponding to the hierarchy of its mechanical structure. Table 1 shows the substructures and algorithms used at different levels in the Autumn 2010 alignment. The numbers in the DoF column represent the product of the number of substructures and the allowed degrees of freedom for each. For example, at Level 2, the Pixel half shells were allowed all three rotations and three translations, while the endcaps only two translations and one rotation \cite{AlignPaper}.
In the latest alignment, discussed in Sec. 4, the Global $\chi^{2}$ was used at all levels but the wire-by-wire alignment of the TRT (approximately 700,000 DoF). The latter used the Local $\chi^{2}$ approach due to computational restrictions.
 
\begin{table}
\caption{Overview of the alignment levels for the Autumn 2010 alignment as described in \cite{AlignPaper}.}
\label{tab:1}       
\begin{tabular}{llclcl}
\hline\noalign{\smallskip}
 Level & Structures & \# DoF & Method\\
\noalign{\smallskip}\hline\noalign{\smallskip}
	& PIX: whole & &\\
Level 1 & SCT: barrel + 2 endcaps & 41 & Global $\chi^2$\\
	& TRT: barrel + 2 endcaps & &\\
\noalign{\smallskip}\hline\noalign{\smallskip}
		& PIX: half shells + disks & &\\
Level 2 & SCT: layers + disks & 852& Global $\chi^2$\\
	& TRT: modules + wheels & &\\
\noalign{\smallskip}\hline\noalign{\smallskip}
	& PIX: modules & &\\
Level 3 & SCT: modules & 722104 & Local $\chi^2$\\
	& TRT: wires & &\\
	\noalign{\smallskip}\hline
\end{tabular}
\end{table}

\section{Alignment performance}
\begin{figure}
\resizebox{0.9\columnwidth}{!}{\includegraphics{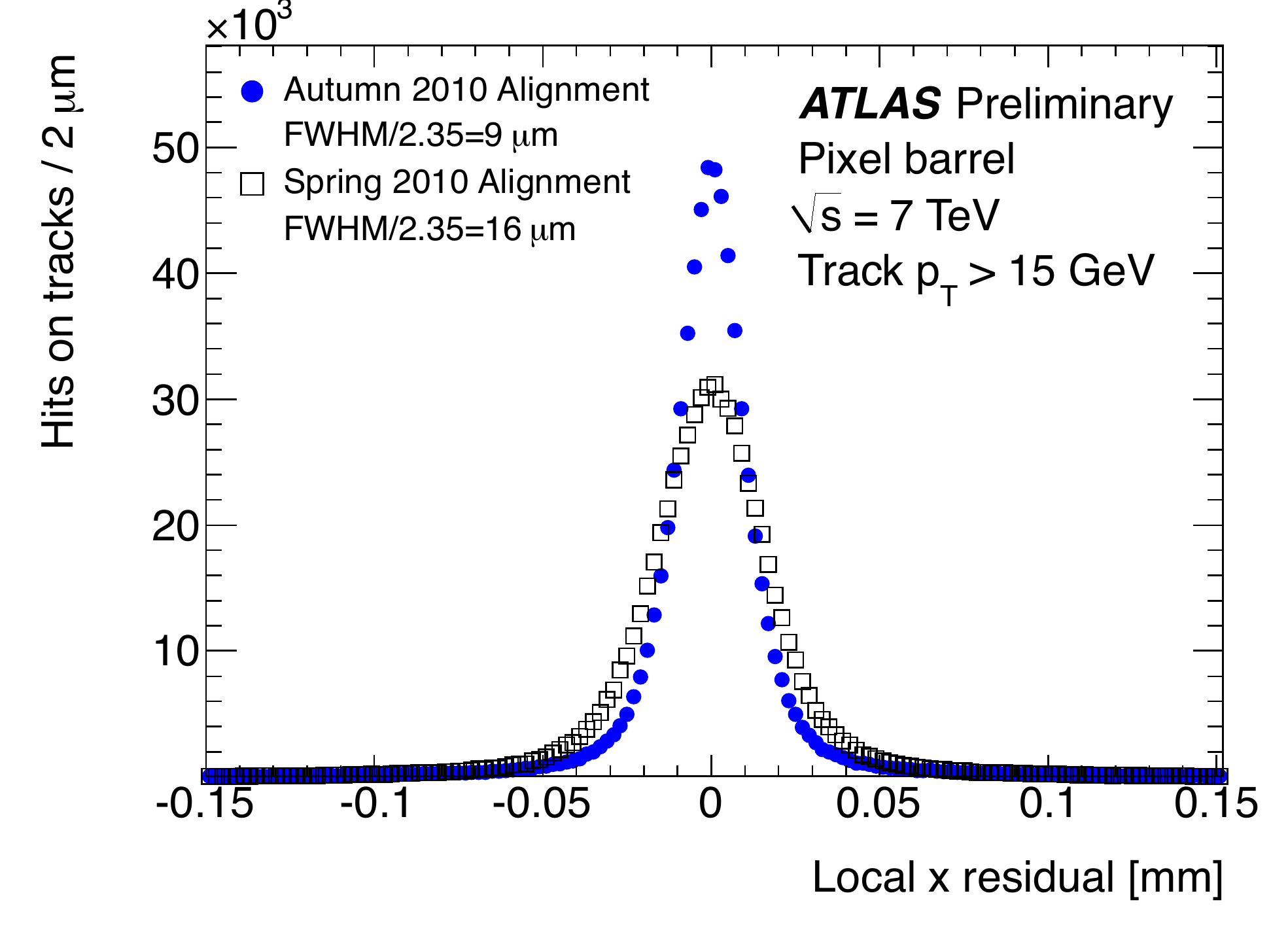}}
\caption{Improvement in resolution in the direction of highest granularity of the Pixel modules after the Autumn 2010 alignment.}
\label{fig:2}       
\end{figure}

\begin{figure}
\resizebox{0.9\columnwidth}{!}{\includegraphics{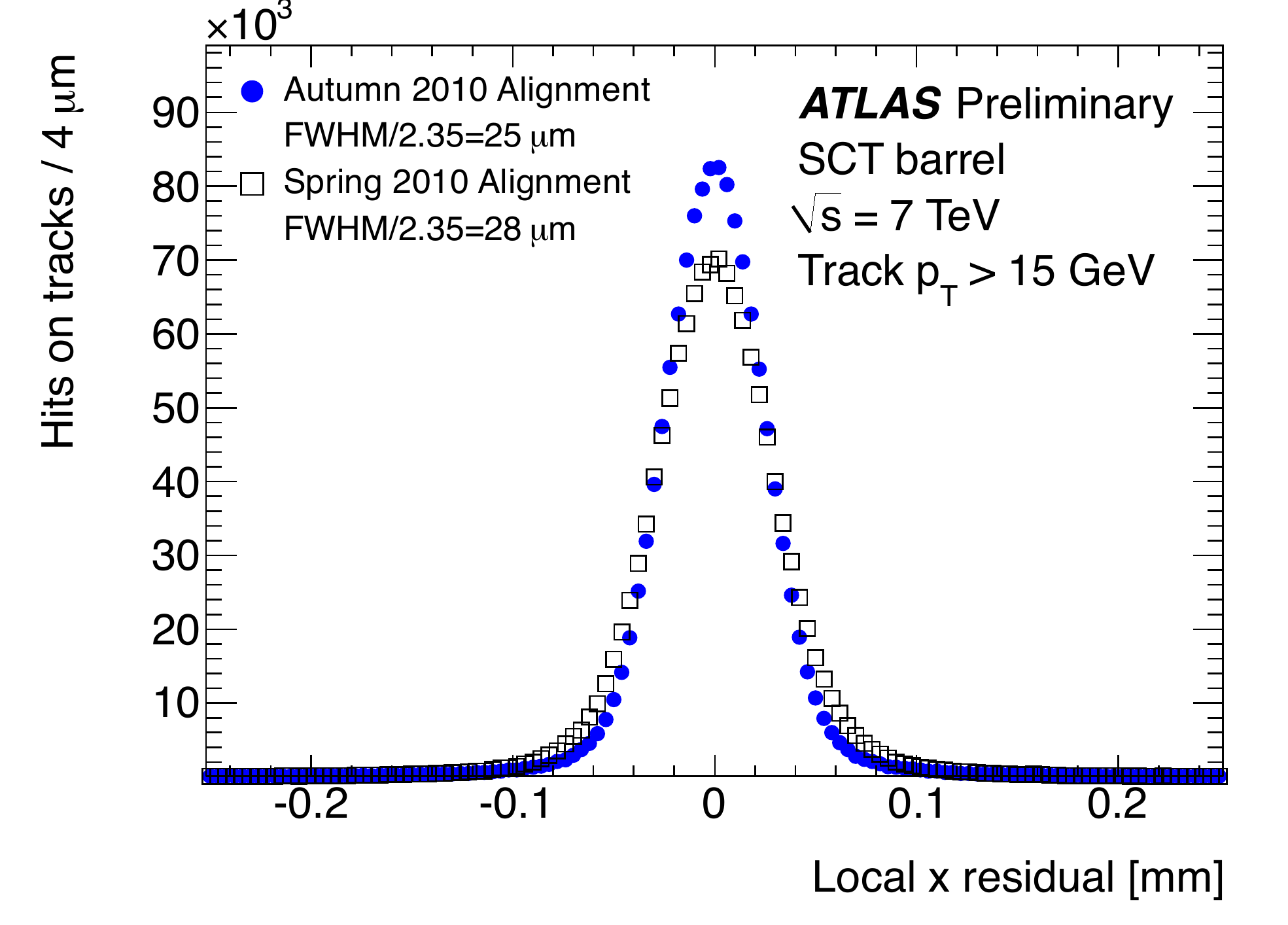}}
\caption{Improvement in resolution in the direction of highest granularity of the SCT modules after the Autumn 2010 alignment.}
\label{fig:3}       
\end{figure}

\begin{figure}
\resizebox{0.9\columnwidth}{!}{\includegraphics{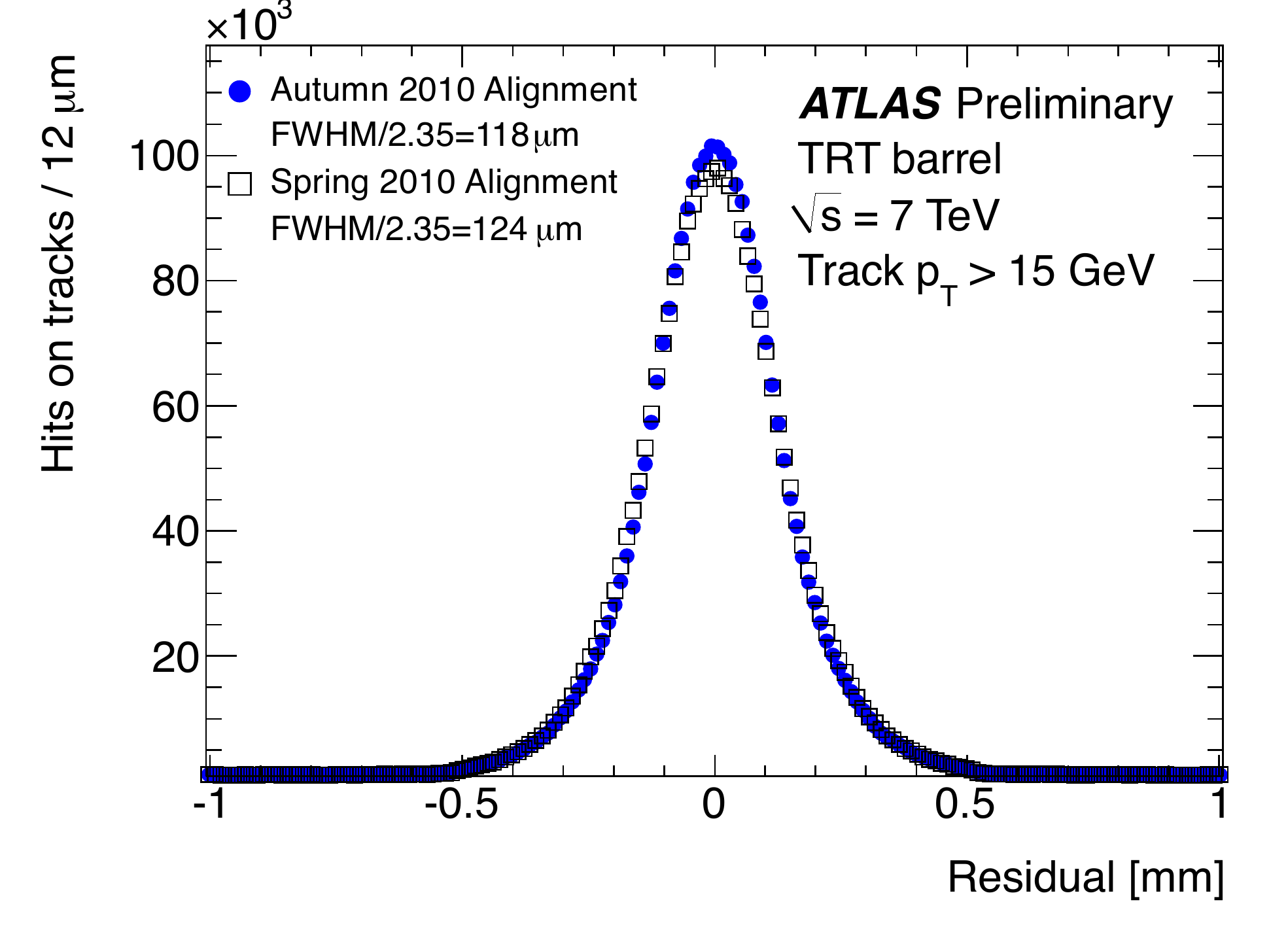}}
\caption{Improvement in resolution in the direction transverse to the TRT wires after the Autumn 2010 alignment.}
\label{fig:4}       
\end{figure}

The Autumn 2010 alignment was the first to use 7 TeV collision data in addition to pixel module wafer deformation input from the production survey. It was also the first wire-by-wire TRT alignment. The impact of these improvements is evident in the reduced width of the residual distributions in the barrel sections of all sub-detectors, the Pixel, SCT and TRT, shown in Figs. \ref{fig:2}, \ref{fig:3} and \ref{fig:4}, respectively. Similar trends are observed in the endcap regions, where the large track statistics used in this alignment were particularly advantageous.\cite{AlignPaper}

\section{Weak modes and constrained alignment}
\begin{figure}
\resizebox{0.9\columnwidth}{!}{\includegraphics{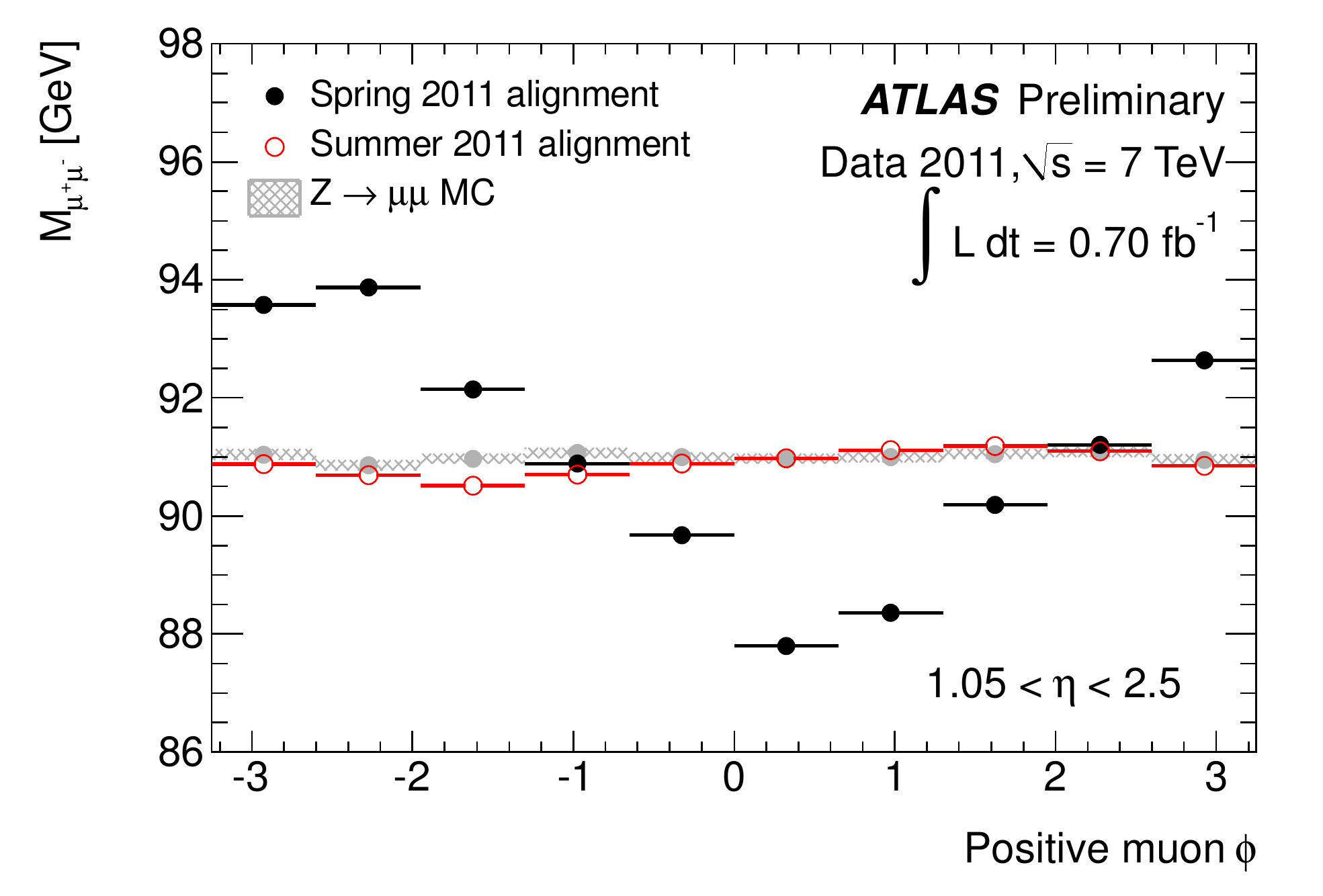}}
\caption{An indication of a weak mode misalignment present in the endcap A region (black) and the effect of using constrained alignment to remove it (red).}
\label{fig:5}       
\end{figure}

\begin{figure}
\resizebox{0.9\columnwidth}{!}{\includegraphics{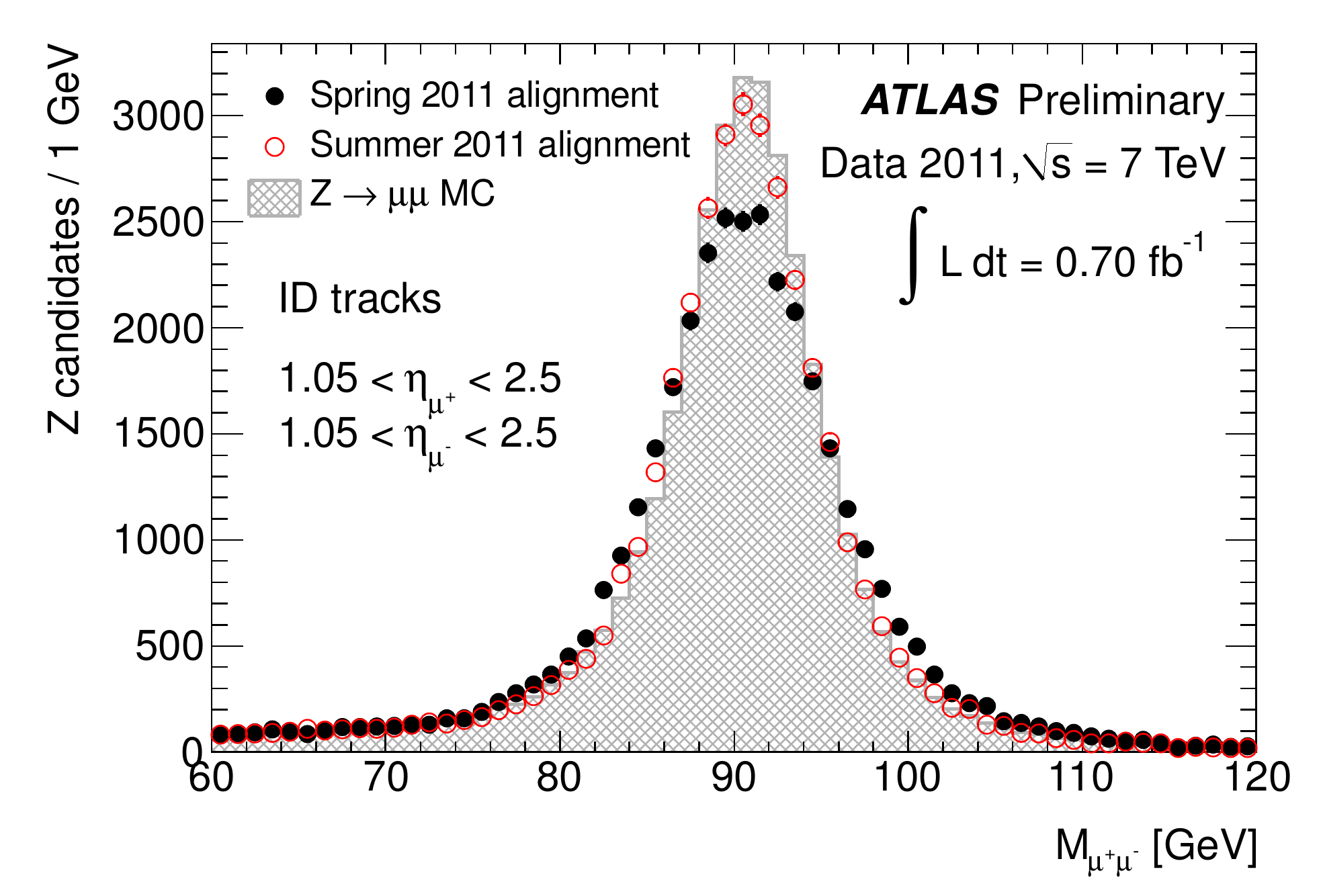}}
\caption{Improvement in the Z mass resolution due to using constrained alignment to remove weak mode misalignment in the endcap A region}
\label{fig:6}       
\end{figure}

There exist systematic detector deformations that cannot be detected using the outlined approach as they retain the helical form of tracks at the expense of biasing the track parameters. They are commonly referred to as ``weak modes'' and can be identified by examining the kinematics of resonance decays such as $Z\rightarrow\mu\mu$, $J/\Psi\rightarrow\mu\mu$ and $K_{S}\rightarrow \pi\pi$. The bias introduced in the track momenta by such misalignments violates the symmetries inherent in these decays and thereby results in unexpected dependences of the reconstructed invariant mass on various kinematic observables. A striking example is the dependence of the Z invariant mass on the $\phi$ track parameter of the positive muon, see Fig. \ref{fig:5}. The approach to correct such misalignments is to constrain some parameters during the alignment, thereby, minimizing the possibility of retaining biases. Some examples of useful constraints are: momentum measurements by the Muon Spectrometer, vertex position constraint and the calorimeter derived constraint. The calorimeter derived constraint, or $E/p$ constraint, uses the fact that the calorimeter response for positrons and electrons should be the same. Differences between the ratio of energy to momentum measurement between electrons and positrons in $Z\rightarrow ee$ or $W\rightarrow e\nu$ decays can then be attributed to mismeasurement of the momentum in the tracker and used to obtain corrections to the reconstructed track momenta in bins of azimuthal angle and pseudorapidity. During alignment, the track momenta are then constrained to the corrected value. The $E/p$ correction has resulted in the latest significant improvement in the alignment as evidenced by the increase in the $Z$ invariant mass resolution shown in Fig. \ref{fig:6}. 

\section{Run-by-run alignment monitoring}
\begin{figure}
\resizebox{0.9\columnwidth}{!}{\includegraphics{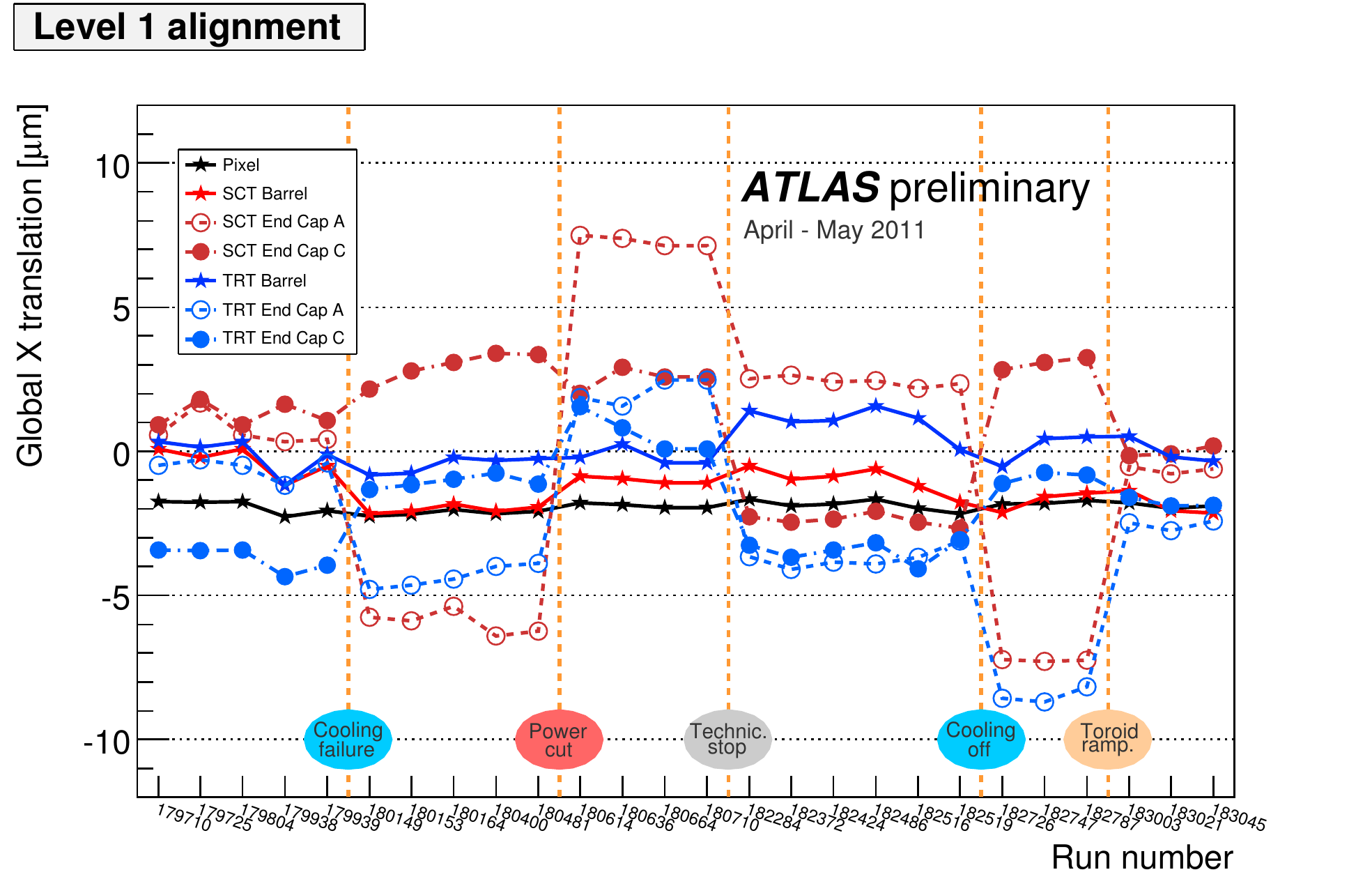}}
\caption{Changes in the alignment constants on a run-by-run basis. Note the impact of incidents such as magnet ramping and cooling interruptions.}
\label{fig:7}       
\end{figure}

Significant changes of the detector alignment occur due to external factors. Some of the identified causes include temperature changes and magnet ramping. The time-ordered global shifts of selected substructures in the direction transverse to the beam pipe are shown in Fig. \ref{fig:7}. The largest changes observed are less than 10 $\mu$m. To monitor and better understand this behavior,  the Level 1 alignment constants are now recomputed on a run-by-run basis.

Additionally, as resonances have been shown to be a powerful probe in uncovering weak-mode misalignments and, thereby, momentum biases, plots of the reconstructed mass as a function of various kinematic variables and the mass itself (as in Figs. \ref{fig:5} and \ref{fig:6}) are also produced automatically for every run as a part of the ATLAS data quality monitoring.

\section{Conclusion and outlook}
The current implementation of the alignment procedure has been shown to be effective and well suited for the challenges posed by the alignment of the ATLAS ID. The next step is to evaluate the systematics caused by residual misalignments. It has already been seen that resonances are a powerful handle for tackling this problem and ongoing studies will soon provide quantitative measures of any remaining biases.

\end{document}